\documentstyle [12pt]{article}
\title{
Progress in Crystal Extraction and Collimation
       }
\author{
A.G.Afonin, \underbar{V.M.Biryukov}, V.T.Baranov, V.N.Chepegin,\\
Y.Chesnokov, V.Kotov, V.Terekhov,
E.Troyanov \\
{\em\small IHEP-Protvino, Russia}\\
V.Guidi, G.Martinelli, M.Stefancich, D.Vincenzi,
{\em\small Ferrara Un., Italy}; \\
Yu.Ivanov,  {\em\small PNPI, Russia};
D.Trbojevic,  {\em\small BNL, USA} \\
W.Scandale,  {\em\small CERN};
M.Breese,  {\em\small Surrey Un., UK}
}
\date{Presented at HEACC (Tsukuba, March 25-30, 2001)}
\begin{document}
\maketitle

\begin{abstract}
\normalsize
Recent IHEP Protvino experiments show efficiencies of crystal-assisted slow
extraction and collimation of 85.3$\pm$2.8\%, at the intensities of the
channeled beam on the order of 10$^{12}$ proton per spill
of $\sim$2 s duration.
The obtained experimental data well
follows the theory predictions.  We compare the measurements against theory
and outline the theoretical potential for further improvement in the
efficiency of the technique. This success is important for the efficient use
of IHEP accelerator and for implementation of crystal-assisted collimation
at RHIC and slow extraction from AGS onto E952, now in preparation.
Future applications, spanning in the energy from order of 1 GeV (scraping in
SNS, slow extraction from COSY and medical accelerators)
to order of 1 TeV and beyond (scraping in
Tevatron, LHC, VLHC), can benefit from these studies.
\end{abstract}

\section{Introduction}

Two major applications of crystal channeling in modern hadron
accelerators are slow extraction and halo collimation
(see e.g. \cite{book} and refs therein).
The benefits of crystal extraction are fourfold.
In hadron colliders this mode of extraction can be made compatible
with the colliding mode of operation.
The time structure of the extracted beam is practically flat,
since the extraction mechanism is resonance-free.
The size of the extracted beam is smaller.
Finally polarized beams can be extracted without
detrimental effects on the polarization.
The benefits of crystal-assisted scraping we discuss in the next section.

These applications can be exploited in a broad range of energies,
from sub-GeV cases (i.e. for medical accelerators) to multi-TeV machines
(for high-energy research). Indeed,
several projects are in progress to investigate them.
Crystal collimation is being studied
at RHIC (100-250 GeV) \cite{rhic},
for the Tevatron (1000 GeV) \cite{tev}
and the LHC \cite{lhc},
for the Spallation Neutron Source (1 GeV) \cite{nuria},
whilst
crystal-assisted slow extraction is considered
for COSY (1-2 GeV)
and AGS (25 GeV) \cite{woody}.

\section{Crystal as a Scraper}

Classic two-stage collimation system for loss localisation
in accelerators typically uses a small scattering target as a primary
element and a bulk absorber as secondary element \cite{jeann}.
The role of the primary element is to give a substantial angular kick
to the incoming particles in order to increase the impact parameter
on the secondary element, which is generally placed in the optimum position
to intercept transverse or longitudinal beam halo.

Naturally, an amorphous target scatters particles in all possible directions.
Ideally, one would prefer a "{\em smart target}" that kicks all particles in
only one direction: for instance, only in radial plane, only outward,
and only into the preferred angular range corresponding to the center of
absorber (to exclude escapes).
Bent crystal is the first idea for such a smart target:
it traps particles and conveys them into the desired
direction.
In physics language, we replace the scattering on single atoms
of amorphous target by the coherent scattering on atomic planes of
aligned monocrystal.

\section{Channeling Efficiency}

It's been long argued theoretically\cite{theory}
that a breakthrough in crystal efficiency can be due to
multiple character of particle encounters with a crystal
installed in a circulating beam.
To clarify this mechanism an extraction experiment was started at IHEP Protvino
at the end of 1997 (see Ref.\cite{pac99,epac2000}
and refs therein).

In the last two years, we demonstrated crystal channeling with
50\% efficiency.
We also showed that these crystals could
be efficiently used as primary collimators, thereby reducing
by a factor two the
radiation level measured downstream of the collimation region of U-70
\cite{pac99,epac2000}.
To continue our investigations,
we installed and tested in U-70 ring several new crystals
produced by different manufacturers with a new shape.
The azimuthal length of the Si(111) crystals was only 1.8-4.0 mm,
bending angle 0.8 to 1.5 mrad.
The advantages of "new-generation" crystals are threefold:
(a) they can be made shorter than a usual bulk crystal,
(b) they have no straight ends, since the bending mechanism is continuous,
and (c) they have no amorphous material close to the beam.
The new technology allows us to control precisely the crystal
length and bending radius.

Two crystals were assembled in Protvino:
one $2$ mm long was bent by 0.9 mrad, the other $4$ mm
long was bent by 1.5 mrad.
The third crystal $1.8$ mm long bent by 0.8 mrad was built and polished
in the University of Ferrara (Italy).
The two Russian crystals were used in extraction mode,
whilst the Italian one was
tested as a primary collimator.
The three crystals were exposed to 70 GeV proton beams and
used to channel and extract halo particles.

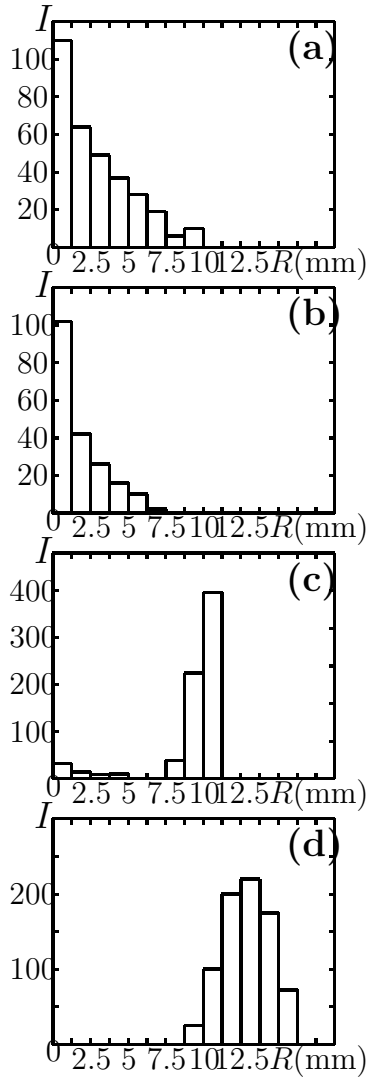
\begin{figure}[htb]
\begin{center}
\setlength{\unitlength}{.25mm}
\begin{picture}(105,160)(20,-40)
\thicklines

\put(0,0) {\line(1,0){150}}
\put(0,0) {\line(0,1){120}}
\put(0,120) {\line(1,0){150}}
\put(150,0){\line(0,1){120}}

\multiput(0,0)(10,0){15}{\line(0,1){3}}
\multiput(0,120)(10,0){15}{\line(0,-1){3}}
\multiput(0,0)(0,20){7}{\line(1,0){3}}
\multiput(150,0)(0,20){7}{\line(-1,0){3}}

\put(0,-9){\makebox(1,1)[b]{0}}
\put(20,-13){\makebox(1,1)[b]{2.5}}
\put(40,-13){\makebox(1,1)[b]{5}}
\put(60,-13){\makebox(1,1)[b]{7.5}}
\put(80,-13){\makebox(1,1)[b]{10}}
\put(100,-13){\makebox(1,1)[b]{12.5}}
\put(-23,100){\makebox(1,.5)[l]{100}}
\put(-19,80){\makebox(1,.5)[l]{80}}
\put(-19,60){\makebox(1,.5)[l]{60}}
\put(-19,40){\makebox(1,.5)[l]{40}}
\put(-19,20){\makebox(1,.5)[l]{20}}

\put(-10,115){\large $I$}
\put(115,-13){$R$(mm)}
\put(124,100){\large\bf (a)}

\linethickness{.3mm}
\put( 70,10)  {\line(1,0){10}}
\put( 60,6)  {\line(1,0){10}}
\put( 50,19)  {\line(1,0){10}}
\put( 40,28)  {\line(1,0){10}}
\put( 30,37)  {\line(1,0){10}}
\put( 20,49)  {\line(1,0){10}}
\put( 10,64)  {\line(1,0){10}}
\put( 0,110)  {\line(1,0){10}}
\put( 80,10)  {\line(0,-1){10}}
\put( 70,10)  {\line(0,-1){10}}
\put( 70,6)  {\line(0,-1){6}}
\put( 60,19)  {\line(0,-1){19}}
\put( 50,28)  {\line(0,-1){28}}
\put( 40,37)  {\line(0,-1){37}}
\put( 30,49)  {\line(0,-1){49}}
\put( 20,64)  {\line(0,-1){64}}
\put( 10,110)  {\line(0,-1){110}}
\end{picture}

\begin{picture}(105,140)(20,-40)
\thicklines

\put(0,0) {\line(1,0){150}}
\put(0,0) {\line(0,1){120}}
\put(0,120) {\line(1,0){150}}
\put(150,0){\line(0,1){120}}

\multiput(0,0)(10,0){15}{\line(0,1){3}}
\multiput(0,120)(10,0){15}{\line(0,-1){3}}
\multiput(0,0)(0,20){7}{\line(1,0){3}}
\multiput(150,0)(0,20){7}{\line(-1,0){3}}

\put(0,-9){\makebox(1,1)[b]{0}}
\put(20,-13){\makebox(1,1)[b]{2.5}}
\put(40,-13){\makebox(1,1)[b]{5}}
\put(60,-13){\makebox(1,1)[b]{7.5}}
\put(80,-13){\makebox(1,1)[b]{10}}
\put(100,-13){\makebox(1,1)[b]{12.5}}
\put(-23,100){\makebox(1,.5)[l]{100}}
\put(-19,80){\makebox(1,.5)[l]{80}}
\put(-19,60){\makebox(1,.5)[l]{60}}
\put(-19,40){\makebox(1,.5)[l]{40}}
\put(-19,20){\makebox(1,.5)[l]{20}}

\put(-10,115){\large $I$}
\put(115,-13){$R$(mm)}
\put(124,100){\large\bf (b)}

\linethickness{.3mm}
\put( 50,2)  {\line(1,0){10}}
\put( 40,10)  {\line(1,0){10}}
\put( 30,16)  {\line(1,0){10}}
\put( 20,26)  {\line(1,0){10}}
\put( 10,42)  {\line(1,0){10}}
\put( 0,102)  {\line(1,0){10}}
\put( 60,2)  {\line(0,-1){2}}
\put( 50,10)  {\line(0,-1){10}}
\put( 40,16)  {\line(0,-1){16}}
\put( 30,26)  {\line(0,-1){26}}
\put( 20,42)  {\line(0,-1){42}}
\put( 10,102)  {\line(0,-1){102}}
\end{picture}

\begin{picture}(105,140)(20,-40)
\thicklines

\put(0,0) {\line(1,0){150}}
\put(0,0) {\line(0,1){120}}
\put(0,120) {\line(1,0){150}}
\put(150,0){\line(0,1){120}}

\multiput(0,0)(10,0){15}{\line(0,1){3}}
\multiput(0,120)(10,0){15}{\line(0,-1){3}}
\multiput(0,0)(0,25){5}{\line(1,0){3}}
\multiput(150,0)(0,20){7}{\line(-1,0){3}}

\put(0,-9){\makebox(1,1)[b]{0}}
\put(20,-13){\makebox(1,1)[b]{2.5}}
\put(40,-13){\makebox(1,1)[b]{5}}
\put(60,-13){\makebox(1,1)[b]{7.5}}
\put(80,-13){\makebox(1,1)[b]{10}}
\put(100,-13){\makebox(1,1)[b]{12.5}}
\put(-23,100){\makebox(1,.5)[l]{400}}
\put(-23,75){\makebox(1,.5)[l]{300}}
\put(-23,50){\makebox(1,.5)[l]{200}}
\put(-23,25){\makebox(1,.5)[l]{100}}

\put(-10,115){\large $I$}
\put(115,-13){$R$(mm)}
\put(124,100){\large\bf (c)}

\linethickness{.3mm}
\put(80,99)  {\line(1,0){10}}
\put(70,56)  {\line(1,0){10}}
\put( 60,9.6)  {\line(1,0){10}}
\put( 50,0)  {\line(1,0){10}}
\put( 40,0)  {\line(1,0){10}}
\put( 30,2.4)  {\line(1,0){10}}
\put( 20,2)  {\line(1,0){10}}
\put( 10,3.6)  {\line(1,0){10}}
\put( 0,8)  {\line(1,0){10}}
\put( 90,99)  {\line(0,-1){99}}
\put( 80,99)  {\line(0,-1){99}}
\put( 70,56)  {\line(0,-1){56}}
\put( 60,9.6)  {\line(0,-1){9.6}}
\put( 40,2.4)  {\line(0,-1){2.4}}
\put( 30,2.4)  {\line(0,-1){2.4}}
\put( 20,3.6)  {\line(0,-1){3.6}}
\put( 10,8)  {\line(0,-1){8}}
\end{picture}

\begin{picture}(105,110)(20,-10)
\thicklines

\put(0,0) {\line(1,0){150}}
\put(0,0) {\line(0,1){120}}
\put(0,120) {\line(1,0){150}}
\put(150,0){\line(0,1){120}}

\multiput(0,0)(10,0){15}{\line(0,1){3}}
\multiput(0,120)(10,0){15}{\line(0,-1){3}}
\multiput(0,0)(0,20){7}{\line(1,0){3}}
\multiput(150,0)(0,20){7}{\line(-1,0){3}}

\put(0,-9){\makebox(1,1)[b]{0}}
\put(20,-13){\makebox(1,1)[b]{2.5}}
\put(40,-13){\makebox(1,1)[b]{5}}
\put(60,-13){\makebox(1,1)[b]{7.5}}
\put(80,-13){\makebox(1,1)[b]{10}}
\put(100,-13){\makebox(1,1)[b]{12.5}}
\put(-23,80){\makebox(1,.5)[l]{200}}
\put(-23,40){\makebox(1,.5)[l]{100}}

\put(-10,115){\large $I$}
\put(115,-13){$R$(mm)}
\put(124,100){\large\bf (d)}

\linethickness{.3mm}
\put(120,29)  {\line(1,0){10}}
\put(110,70)  {\line(1,0){10}}
\put(100,88)  {\line(1,0){10}}
\put( 90,80)  {\line(1,0){10}}
\put( 80,40)  {\line(1,0){10}}
\put( 70,10)  {\line(1,0){10}}
\put(130,29)  {\line(0,-1){29}}
\put(120,70)  {\line(0,-1){70}}
\put(110,88)  {\line(0,-1){88}}
\put(100,88)  {\line(0,-1){88}}
\put( 90,80)  {\line(0,-1){80}}
\put( 80,40)  {\line(0,-1){40}}
\put( 70,10)  {\line(0,-1){10}}
\end{picture}

\end{center}
\caption{
Beam profiles measured at the collimator entry face:
(a) crystal out;
(b) crystal in, but misaligned;
(c) crystal in the beam, aligned;
(d) beam kicked by magnet.
}
  \label{col2}
\end{figure}

Fig.1 illustrates
the beneficial effect of crystals
when used as primary collimators.
We present beam profiles in the radial plane
downstream of the crystal,
recorded with the profile-meter of Ref.\cite{pac99}. The coordinate
$R$ represents the radial displacement referred
to the collimator edge.
Four cases are reported.
In first one, an amorphous collimator is used as primary target
whilst the close-by crystal is kept outside of the beam envelope.
As expected, the beam profile is peaked at the collimator edge (Fig. 1(a)).
In the second case (Fig. 1(b)) the
crystal is used as the primary scraper,
whilst the amorphous target is retracted.
No care is taken to
align crystal with respect to the beam direction,
hence its action on the incoming protons is very similar to that
of an amorphous target.
When properly aligned (see Fig. 1(c)), the crystal channels most
of the incoming beam and displaces their distribution by
about $10$ mm inside the collimator edge.
In the last case (see Fig. 1(d)),
the beam is simply kicked by a magnet towards the secondary collimator,
whilst the primary target is retracted.

\begin{figure}[thb]
\begin{center}
\setlength{\unitlength}{.5mm}
\begin{picture}(110,105)(15,0)
\thicklines
\linethickness{.3mm}
\put(   8,91)  {\circle{2.5}}
\put(  10,92)  {\circle{2.5}}
\put(  12,92)  {\circle{2.5}}
\put(  14,92)  {\circle{2.5}}
\put(  16,92)  {\circle{2.5}}
\put(  20,90)  {\circle{2.5}}
\put(  24,89)  {\circle{2.5}}
\put(  30,87)  {\circle{2.5}}
\put(  40,83)  {\circle{2.5}}
\put(  50,79)  {\circle{2.5}}
\put(  60,75)  {\circle{2.5}}

\put(  78,67)  {\LARGE $\star$}
\put(  34,84)  {\LARGE $\star$}
\put(  38,84)  {\LARGE $\star$}

\put( 138,12)  {$\otimes$}
\put(  58,62)  {$\Box$}
\put(  98,48)  {$\Box$}

\put(0,0) {\line(1,0){150}}
\put(0,0) {\line(0,1){100}}
\put(0,100) {\line(1,0){150}}
\put(150,0){\line(0,1){100}}

\multiput(4,0)(4,0){37}{\line(0,1){2}}
\multiput(20,0)(20,0){7}{\line(0,1){4}}
\multiput(4,100)(4,0){37}{\line(0,-1){2}}
\multiput(20,100)(20,0){7}{\line(0,-1){4}}
\multiput(0,20)(0,20){4}{\line(1,0){3}}
\multiput(0,4)(0,4){25}{\line(1,0){1.4}}
\multiput(150,20)(0,20){4}{\line(-1,0){3}}
\multiput(150,4)(0,4){25}{\line(-1,0){1.4}}

\put(0,-7){\makebox(1,1)[b]{0}}
\put(20,-7){\makebox(1,1)[b]{1}}
\put(40,-7){\makebox(1,1)[b]{2}}
\put(60,-7){\makebox(1,1)[b]{3}}
\put(80,-7){\makebox(1,1)[b]{4}}
\put(100,-7){\makebox(1,1)[b]{5}}
\put(120,-7){\makebox(1,1)[b]{6}}
\put(140,-7){\makebox(1,1)[b]{7}}
\put(-11,20){\makebox(1,.5)[l]{20}}
\put(-11,40){\makebox(1,.5)[l]{40}}
\put(-11,60){\makebox(1,.5)[l]{60}}
\put(-11,80){\makebox(1,.5)[l]{80}}
\put(-13,100){\makebox(1,.5)[l]{100}}

\put(10,105){ Extraction efficiency (\%)}
\put(60,-14){ Crystal length (mm)}

\end{picture}
\end{center}
\caption{ %\large
Crystal extraction efficiency as measured for 70-GeV protons.
Recent results ($\star$, strips 1.8, 2.0, and 4 mm),
1999-2000 ($\Box$, O-shaped crystals 3 and 5 mm),
and 1997 ($\otimes$, strip 7 mm).
Also shown (o) is Monte Carlo prediction [7]
for a perfect crystal.
}
  \label{mcs4}
\end{figure}
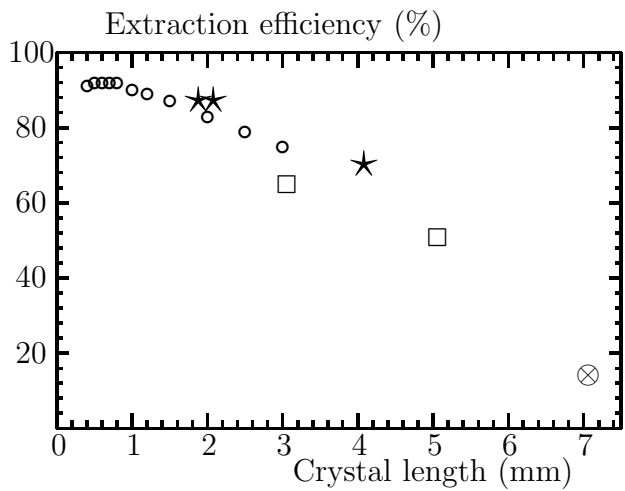

The channeling efficiency is given by
the ratio of the extracted beam intensity, as measured in
the external beam line, to all the beam loss, as measured in the entire ring;
see the diagnostics part of the experiment described
in refs. \cite{pac99,epac2000}.
We obtained very high channeling efficiencies in {\em each} of the
three new crystals: namely, both the 1.8 and 2 mm long crystals reached
85\% efficiency, whilst the 4 mm long crystal
reached 68\% efficiency.
In Fig.\ref{mcs4} we plot the expected (the prediction published in
\cite{epac2000}) and the measured channeling
efficiencies together with data relative to an old O-shaped crystal.
The agreement between measurements and simulations is excellent.
Fig.\ref{mcs4} shows the theoretical potential for channeling efficiencies
of 90-95\% when we manage a crystal deflector with the size optimal
for our set-up.

These unprecedented results were indeed obtained in a steady manner
over many runs. In particular, the 2 mm long crystal was regularly
functioning to extract beams with a channelling efficiency of 85.3$\pm$2.8\%.

\section{High-intensity tests}

Beside the channeling efficiency, also important are
standing a high beam intensity and
crystal lifetime.
Crystals located in the region upstream of the U-70 cleaning area were
irradiated with the entire circulating beam, spilled out in rather
short time durations to simulate very dense halo collimation.
We can measure precisely
the beam intensity intentionally damped into
the crystal. However, we can only estimate with computer simulations
the total amount of particle hits during a spill, since
unchanneled protons are simply scattered and
may continue to circulate in the ring hitting the crystal many times.
The number of hits per primary particle can vary from a few
to more than hundred.
Such analysis has shown that
our crystals were irradiated up to
$2 \times 10^{14}$ particles
per spill of $\sim$1 s duration.
When averaged over machine cycles,
the irradiation rate was as high as
$2 \times 10^{13}$ proton hits/s.

Notice that this irradiation rate
already {\em exceeds} the expected beam
loss rate at the Spallation Neutron Source. Indeed,
the SNS Accumulator Ring should generate a 1 GeV
proton flux
of 60$\times$2$\times$10$^{14}$ per second. At the expected rate of
beam loss of 0.1\% the halo flux will be
1.2$\times$10$^{13}$ protons/s.
Several crystals in use in U-70 have been exposed to high intensity beams
for months.
After the irradiation of $\sim$10$^{20}$p/cm$^2$
the initial channelling efficiency was practically unaffected.

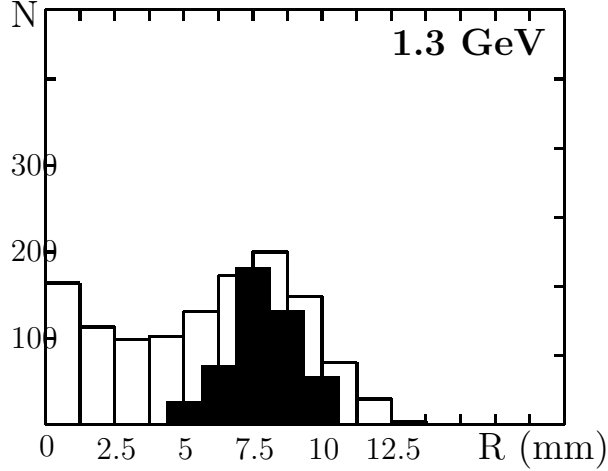
\begin{figure}[h]
\begin{center}
\setlength{\unitlength}{.46mm}
\begin{picture}(105,120)(20,0)
\thicklines

\put(0,0) {\line(1,0){150}}
\put(0,0) {\line(0,1){120}}
\put(0,120) {\line(1,0){150}}
\put(150,0){\line(0,1){120}}

\multiput(0,0)(10,0){15}{\line(0,1){3}}
\multiput(0,120)(10,0){15}{\line(0,-1){3}}
\multiput(0,0)(0,25){5}{\line(1,0){3}}
\multiput(150,0)(0,20){7}{\line(-1,0){3}}

\put(0,-9){\makebox(1,1)[b]{0}}
\put(20,-10){\makebox(1,1)[b]{2.5}}
\put(40,-10){\makebox(1,1)[b]{5}}
\put(60,-10){\makebox(1,1)[b]{7.5}}
\put(80,-10){\makebox(1,1)[b]{10}}
\put(100,-10){\makebox(1,1)[b]{12.5}}
\put(-10,75){\makebox(1,.5)[l]{300}}
\put(-10,50){\makebox(1,.5)[l]{200}}
\put(-10,25){\makebox(1,.5)[l]{100}}

\put(-10,115){\large N}
\put(125,-10){\large R (mm)}
\put(100,107){\large\bf 1.3 GeV}

\linethickness{.3mm}
\put(100,0.9)  {\line(1,0){10}}
\put( 90,7.5)  {\line(1,0){10}}
\put( 80,18)  {\line(1,0){10}}
\put( 70,37)  {\line(1,0){10}}
\put( 60,50)  {\line(1,0){10}}
\put( 50,43.2)  {\line(1,0){10}}
\put( 40,32.7)  {\line(1,0){10}}
\put( 30,25.5)  {\line(1,0){10}}
\put( 20,24.7)  {\line(1,0){10}}
\put( 10,28.3)  {\line(1,0){10}}
\put( 0,41)  {\line(1,0){10}}

\put( 110,0.9)  {\line(0,-1){0.9}}
\put( 100,7.5)  {\line(0,-1){7.5}}
\put( 90,18)  {\line(0,-1){18}}
\put( 80,37)  {\line(0,-1){37}}
\put( 70,50)  {\line(0,-1){50}}
\put( 60,50)  {\line(0,-1){50}}
\put( 50,43.2)  {\line(0,-1){43.2}}
\put( 40,32.7)  {\line(0,-1){32.7}}
\put( 30,25.5)  {\line(0,-1){25.5}}
\put( 20,28.3)  {\line(0,-1){28.3}}
\put( 10,41)  {\line(0,-1){41}}

\linethickness{4.5mm}
\put( 80,0)  {\line(0,1){14}  }
\put( 70,0)  {\line(0,1){33.2}}
\put( 60,0)  {\line(0,1){45.6}}
\put( 50,0)  {\line(0,1){17.2}}
\put( 40,0)  {\line(0,1){6.7} }
\end{picture}

\end{center}
\caption{
Beam profile as measured on the collimator entry face
with 1.3 GeV protons.
In black is shown the simulated profile of channeled protons.
}
  \label{colg}
\end{figure}

\section{Collimation at 1.3 GeV}

On the same location in U-70
with the same 1.8-mm crystal of Si(111)
positioned $\sim$20 m upstream of the ring collimator,
we have repeated the crystal collimation experiment
at the injection flattop of U-70,
proton kinetic energy of 1.3 GeV.
With the crystal aligned to the incoming halo particles,
the radial beam profile at the collimator entry face
showed a significant channeled peak far from the edge, Fig.\ref{colg}.

The expected width of the channeled peak
is about 5 bins in the profile
of Fig.\ref{colg}, in agreement with observations.
About half of the protons intercepted by the collimator jaw,
have been channeled there by a crystal; i.e., crystal
doubled the amount of particles intercepted by the jaw.
As only part (about 34\%) of all particles scattered off the crystal
have reached the jaw, we estimate the
crystal deflection efficiency as 15-20\%.
The observed
figure of efficiency could be well reproduced in computer simulations.
This figure is orders of magnitude higher than prevous world data
for low-GeV energy range.

It is remarkable that the same crystal was efficiently channeling
both at 70 GeV and at 1.3 GeV,
thus demonstrating to be operational in a very wide energy range.

\section {Conclusion}

The crystal channeling efficiency has reached unprecedented high values
both at top energy and at injection energy.
The same 2 mm long crystal was used to channel 70 GeV protons
with an efficiency of 85.3$\pm$2.8\% during several weeks of operation
and 1.32 GeV protons with an efficiency of 15-20\% during some test runs.
Crystals with a similar design were able to stand radiation doses
over $10^{20}$ proton/cm$^2$
and irradiation rates of $2 \times 10^{14}$ particles
incident on crystal in spills of $\sim$2 s duration
without deterioration of their performances.

The efficiency results well match the figures theoretically expected
for ideal crystals. As simulations show, extraction
and collimation with channeling efficiencies over 90-95\% is feasible.
The obtained high figures provide
a crucial support for the ideas to apply this technique in beam cleaning
systems, for instance in RHIC and Tevatron.
Earlier Tevatron scraping simulations \cite{tev} have shown that crystal
scraper reduces accelerator-related background in CDF and D0
experiments by a factor of $\sim$10.
This year, first experimental data is expected from RHIC
where crystal collimator \cite{rhic} is installed.
The technique presented here is potentially applicable also in LHC for
instance to improve the efficiency of the LHC cleaning system by embedding
bent crystals in the primary collimators\cite{lhc}.

This work was supported by INTAS-CERN grant 132-2000,
RFBR grant 01-02-16229, and by the
"Young researcher Project" of the University of Ferrara.

\end{document}